\documentclass[twocolumn,aps,prl]{revtex4}

\usepackage{graphics}% Include figure files
\usepackage{graphicx}% Include figure files
\usepackage{dcolumn}% Align table columns on decimal point
\usepackage{bm}% bold math
\usepackage{amsmath,amssymb}

\newcommand{\ket}[1]{|#1\rangle}

\newcommand{\de}{\partial}

\newcommand{\eq}[2]{\begin{equation} \label{#1} #2 \end{equation}}

\newcommand{\sgn}{\textrm{sgn}}
\newcommand{\pv}{\mathbb{P}\int_{-\infty}^{+\infty}}

\newcommand{\diag}{\textrm{diag}}
\newcommand{\etal}{{\em et al.}}
\newcommand{\integrale}{\int_{-\infty}^{+\infty}}

\newcommand{\schr}{Schr\"odinger }

\begin{document}

%\draft

% Change notation for vectors, should not be mixed with commutators

\title{Hybrid squeezing of solitonic resonant radiation in photonic crystal fibers}
\author{Truong X. Tran$^{1}$, Katiuscia N. Cassemiro$^{1}$, Christoph S\"oller$^{1}$, Keith J. Blow$^{2}$ and Fabio Biancalana$^{1}$}
\affiliation{$^{1}$Max Planck Institute for the Science of Light, 91058 Erlangen, Germany}
\affiliation{$^{2}$Aston University, Aston Triangle, Birmingham, B4 7ET, UK}
\date{\today}

\begin{abstract}
We report on the existence of a novel kind of squeezing in photonic crystal fibers which is conceptually intermediate between the four-wave mixing induced squeezing, in which all the participant waves are monochromatic waves, and the self-phase modulation induced squeezing for a single pulse in a coherent state. This {\em hybrid squeezing} occurs when an arbitrary short soliton emits quasi-monochromatic resonant radiation near a zero group velocity dispersion point of the fiber. Photons around the resonant frequency become strongly correlated due to the presence of the classical soliton, and a reduction of the quantum noise below the shot noise level is predicted.
\end{abstract}

\maketitle

{\it Introduction --} In quantum mechanics the variance of two conjugate observables must respect a minimum value set by the Heisenberg uncertainty principle, known as standard quantum limit or shot-noise level. Squeezing is the result of strong correlations, generally induced by nonlinear optical processes, such that the variance of a given observable is smaller than the shot noise~\cite{foxbook}. Squeezed light constitutes an important tool for high-precision measurements~\cite{goda}, as well as for applications in quantum cryptography and quantum information processing (see Ref. \cite{obrien} and references therein). Two qualitatively different kinds of squeezing are currently known to take place in optical fibers. The first is due to the process of four-wave mixing (FWM)~\cite{shelbyfwm}, in which an intense continuous wave (CW) pump field generates signal and idler waves at symmetric frequencies around the pump. Energy conservation leads to a strict photon number correlation between the signal and idler twin modes, resulting in squeezing. The second kind of squeezing is due to self-phase modulation (SPM)~\cite{rosenbluh}, in which the spectral components of an intense short pulse (e.g. a soliton~\cite{spaelter}) in a coherent state become correlated. This is different from the FWM-induced squeezing, in that the input field is pulsed and is itself squeezed as opposed to squeezing a vacuum state~\cite{slusher,blow}.

In this paper we theoretically demonstrate that a third kind of squeezing is possible in optical fibers. We investigate the quasi-monochromatic dispersive radiation emitted by optical solitons subjected to perturbations. Evidence of photon correlation is observed, manifesting itself in a noise spectrum of the emitted radiation lower than the standard quantum limit.

{\it Resonant radiation amplitude with Green's function method --}
Our starting point is the generalized nonlinear \schr equation (GNLSE) in the absence of Raman effect:
$i\de_{z}A+D(i\de_{t})A+|A|^{2}A=0$. Here $z$ and $t$ are the dimensionless time and space coordinates, scaled with the input pulse duration $t_{0}$ and the second-order dispersion length $L_{D2}\equiv t_{0}^{2}/|\beta_{2}(\omega_{0})|$, respectively. $A(z,t)$ is the dimensionless electric field envelope in units of $\sqrt{P_{0}}$, with $P_{0}\equiv|\beta_{2}(\omega_{0})|/(\gamma t_{0}^{2})$, $\beta_{2}$ is the group velocity dispersion (GVD) coefficient at the central frequency $\omega_{0}$ of the input pulse, and $\gamma$ is the nonlinear coefficient of the fiber \cite{agrawalbook2}. The linear dispersion operator in the GNLSE is given by $D(i\de_{t})\equiv \frac{1}{2} s \de_{t}^{2} + \sum_{m\geq3}\alpha_{m}[i\de_{t}]^{m}$, where $s=+1$ ($s=-1$) for anomalous (normal) GVD, $\alpha_{m}\equiv\beta_{m}/[m!|\beta_{2}|t^{m-2}_{0}]$, where $\beta_{m}$ is the $m$-th order GVD coefficient.
When no higher-order dispersion (HOD) coefficients are taken into account ($\alpha_{m}=0, \forall m\geq3 $), the GNLSE takes the standard form of the conventional NLSE with the following fundamental soliton solution $A(z,t)=F(t)\exp(iqz)$, with $F(t)\equiv\sqrt{2q}\mathrm{sech}(\sqrt{2q}t)$, where $q$ is the soliton wavenumber \cite{agrawalbook2}.
In absence of HOD, the fundamental soliton is stable and invariant along propagation. When the NLSE is perturbed by HOD terms, solitons propagating according to the GNLSE emit a special kind of quasi-monochromatic dispersive radiation, called {\em resonant} (or {\em Cherenkov}) {\em radiation} (RR), provided that phase-matching between soliton wavenumber and the fiber dispersion is met \cite{akhmediev,biancalanaradiation}.

We now proceed to calculate the amplitude of RR emitted by a soliton in the presence of HOD by using a Green's function method, a procedure that is introduced for the first time in this paper. This will provide an analytical expression for the spatiotemporal evolution of creation and annihilation operators of the RR around the resonant frequency. These will be used to calculate explicitly the quadrature variance, in order to investigate quantitatively the squeezing effect along the fiber. We make the following {\em ansatz} in the GNLSE: $A= \left[F(t)+\hat{g}(z,t)\right]\exp(iqz)$, where $\hat{g}$ ($\hat{g}^{\dagger}$) is the quantum annihilation (creation) operator associated to the radiation amplitude, which is much smaller than the soliton amplitude. By keeping only the linear terms in $\vec{g}\equiv(\hat{g},\hat{g}^{\dagger})^{\rm T}$ we obtain:
\eq{fullEq}{\left[i\hat{\sigma}\de_{z}-q+\frac{1}{2}\de_{t}^{2}-\hat{P}(i\de_{t})\right]\vec{g}= \vec{U}(\vec{F}) - \hat{M}\vec{g},} where
$\hat{\sigma}=\diag(1,-1)$ is the third Pauli matrix. The soliton potential matrix $\hat{M}(t)$ and the HOD perturbation operator $\hat{P}(i\de_{t})$ are defined by (from now on an overbar indicates complex conjugation):
\eq{Mform}{\small \hat{M}(t)=\left[\begin{array}{cc} 2|F(t)|^{2} & F^{2}(t)
\\ \bar{F}^{2}(t) & 2|F(t)|^{2}
\end{array}\right],\hat{P}(i\de_{t})=\left[\begin{array}{cc} P(i\de_{t}) & 0 \\ 0 &
P(-i\de_{t})
\end{array}\right],} where
$P(i\de_{t})\equiv[D_{2}(i\de_{t})-D(i\de_{t})]$, $D_{2}\equiv\frac{1}{2} s \de_{t}^{2}$,
$\vec{U}(\vec{F})\equiv(P(i\de_{t})F,P(-i\de_{t})\bar{F})^{\rm T}=\hat{P}(i\de_{t})\vec{F}$,
with $\vec{F}=(F,\bar{F})^{\rm T}$. Hermitian matrix $\hat{M}$
is the potential felt by radiation $\vec{g}$ due to the soliton, while vector $\vec{U}(\vec{F})$
is the radiation's source term due to HOD perturbations. Note
that because $\hat{M}$ is non-diagonal, field $\hat{g}$ will be coupled to
field $\hat{g}^\dagger$ for $z\neq 0$, even if there is no initial coupling at $z=0$. This feature will eventually be responsible for the squeezing of RR, which is thus indirectly induced by the soliton itself.

We now solve Eq. (\ref{fullEq}) by first solving the following equation for the Green's function $\mathcal{G}$: \eq{greenf1}{\left[i\de_{z}-q+ D(i\de_{t})\right]\mathcal{G}(z,t-t')=\Delta( t-t'),} where $\Delta(t)$ is the Dirac delta function. Once $\mathcal{G}$ is known, the convolution theorem allows one to find the general solution of Eq. (\ref{fullEq}): \eq{gz}{\small \vec{g}(z,t)=\vec{g}_{0}(t)+\integrale
\hat{\mathcal{G}}(z,t-t')\left[\vec{U}(\vec{F}(t'))-\hat{M}(t')\vec{g}(z,t') \right]dt',} where
$\vec{g}_{0}$ is the quantum radiation field at $z=0$ (corresponding to vacuum fluctuations), and we have defined the
$2\times 2$ matrix $\hat{\mathcal{G}}=\diag(\mathcal{G},\bar{\mathcal{G}})$. Note that Eq. (\ref{gz}) is an {\em integral equation}: the unknown radiation field
$\vec{g}(z,t)$ appears on both sides. To solve Eq. (\ref{gz}), one needs first to find Green's function $\mathcal{G}$ by solving Eq. (\ref{greenf1}), and then by using an iterative perturbative procedure it will be possible to find $\vec{g}(z,t)$ with an arbitrary accuracy.

It is possible to rigorously show by using the residue theorem of complex analysis that the $z$-dependent Green's function $\mathcal{G}$ is given by \eq{rx15}{\mathcal{G}(z,t-t')=\frac{1}{2\pi}\pv\frac{e^{-i\omega[t-t']}
\{1-e^{i[-q+D(\omega)]z}\} }{-q+D(\omega)}d\omega,} where $\mathbb{P}$ indicates the Cauchy principal value of the integral, and the integrand is an analytic function.
An analysis of the pole structure indicates that if the soliton emits RR due to perturbations, the resonant frequencies will be solutions of the equation $-q+D(\omega)=0$, where $D(\omega)$ is the Fourier transform of the dispersion operator $D(i\de_{t})$. We assume in the following (without loss of generality) that there will be only one {\em real} root of this polynomial equation, namely $\omega=\delta_{0}$. This occurs easily in the context of the third-order dispersion induced RR emitted by PCFs \cite{biancalanaradiation}.
Moreover, one should note that the integrand in Eq. (\ref{rx15}) contains an exponential with
argument $i[-q+D(\omega)]z$, which can be, generally speaking, a complicated
function of the complex frequency $\omega$. It is therefore not possible to
solve the integral of Eq. (\ref{rx15}) explicitly, unless a specific integration
path satisfying Jordan's lemma \cite{jordanlemma} is found. However, such a path can be found once
an additional (but not excessively restrictive) assumption is made, i.e. that
the function $[-q+D(\omega)]$ can be expanded in a Taylor series around the
only resonance $\delta_{0}$:
\eq{rx15bis}{\small [-q+D(\omega)]\simeq\left[\frac{\de[-q+D(\omega)]}{\de\omega}\right]_{\omega=\delta_{0}}[\omega-\delta_{0}]\equiv
v_{g}^{-1}[\omega-\delta_{0}],} where $v_{g}(\delta_{0})$ is related to the group velocity at the RR frequency. The explicit form of $\hat{\mathcal{G}}=\diag(\mathcal{G},\bar{\mathcal{G}})$
deduced from Eqs. (\ref{rx15}-\ref{rx15bis}) is given by
\eq{explicitform}{\hat{\mathcal{G}}(z,t-t')=
iv_{g}\hat{\sigma}e^{i\hat{\sigma}\delta_{0}[t'-t]}\hat{\eta}_{z}(v_{g},t'-t),}
where $\hat{\eta}_{z}(v_{g},t'-t)=\diag(\eta_{z}(v_{g},t'-t),\eta_{z}(v_{g},t'-t))$, and $\eta_{z}(v_{g},t'-t)\equiv[\sgn(t'-t)-\sgn(t'-t+v_{g}^{-1}z)]/2$ is a useful 'truncation' function that appears often in the calculations, and physically provides the position of the RR emitted by the soliton in the time domain.
We are now in the position to use the convolution theorem Eq. (\ref{gz}) and evaluate the
radiation field $\vec{g}(z,t)$ perturbatively. Let us define a new matrix kernel
$\hat{K}(z;t,t')\equiv-\hat{\mathcal{G}}(z,t)\hat{M}(t')$,
and a 'constant' vector $\vec{G}_{0}(z,t)\equiv\integrale \hat{\mathcal{G}}(z,t-t')\vec{U}(\vec{F}(t'))dt'$, which physically represents the classical background field associated with the soliton tail, source of the RR classical parametric growth. With these new
definitions we can write Eq. (\ref{gz}) in the following
symbolic way: \eq{rx21}{\vec{g}=\vec{g}_{0}+\vec{G}_{0}+\hat{K}\otimes\vec{g},} where the
symbol $\otimes$ denotes the operation of convolution with respect to the time variable.
Using this symbolic notation, one can rearrange the terms of Eq. (\ref{rx21}) to
obtain a formal solution for the amplitude field in terms of a Lippmann-Schwinger equation \cite{taylor}
\eq{rx22}{\vec{g}=\frac{1}{1-\hat{K}\otimes}(\vec{g}_{0}+\vec{G}_{0}).} We can expand the operator $[1-\hat{K}\otimes]^{-1}$ in a series,
$[1-\hat{K}\otimes]^{-1}=1+\hat{K}\otimes+\hat{K}\otimes\hat{K}\otimes+\dots$. The $0$-th order in the perturbation theory is given by
$\vec{g}^{(0)}(z,t)=\vec{g}_{0}(t) + \vec{G}_{0}(z,t)$. At this level the {\em
interaction} between the solitonic potential $\hat{M}$ and the radiation is totally
neglected, the coupling between fields $\hat{g}$ and $\hat{g}^{\dagger}$ in Eqs. (\ref{main1}-\ref{main2}) is not present and squeezing is not possible. Note that, however, the $0$-th order gives reasonable approximate results if one is only interested in the calculation of the classical RR amplitude \cite{biancalanaradiation,akhmediev}.
To the next order in the Born series, one obtains
\eq{rx26}{\vec{g}^{(1)}=[1+\hat{K}\otimes](\vec{g}_{0}+\vec{G}_{0}) = \vec{g}_{0}+\hat{K}\otimes\vec{g}_{0} +  [\vec{G}_{0}+\hat{K}\otimes\vec{G}_{0}],} which already
takes into account the potential matrix $\hat{M}$ hidden in the kernel operator $\hat{K}$.
The term in square brackets in Eq. (\ref{rx26}) represents the influence of HOD acting on the soliton, and is a purely 'classical' quantity. In this paper the main focus is on the quantum noise properties of the RR. Thus we decouple the classical background soliton tail $\vec{G}_{0}$ from the equations, since this does not affect the evolution of the quantum fluctuations around the resonant frequency.

Now we can specify the initial quantum vacuum field at the resonant frequency $\delta_{0}$ (the frequency detuning from the soliton carrier frequency):
$\vec{g}_{0}(t) \equiv(\hat{a}_{0}(t),\hat{a}_{0}^\dagger(t))^{\rm T}$,
where $\hat{a}_{0}^\dagger$ and $\hat{a}_{0}$ are, respectively, the $t$-dependent creation and destruction operators at the input of the fiber ($z=0$), which satisfy the shot noise commutation relation $[\hat{a}_{0}(t),\hat{a}_{0}^{\dagger}(t')]=\Delta(t-t')$. In quantum optics this vacuum fluctuation field is everywhere and always exists even in the absence of classical signals. Due to Eq. (\ref{rx26}), vacuum noise fluctuations interact with the spectrally separated soliton, and acquire a $z$-evolution given by $\vec{g}^{(1)}(z,t) \equiv(\hat{a}(z,t),\hat{a}^{\dagger}(z,t))^{\rm T}$.

From the above definition of the kernel operator $\hat{K}$ we have:
\eq{coupler1}{\hat{K}\otimes\vec{g}_{0}\equiv-\integrale\hat{\mathcal{G}}(z,t-t')\hat{M}(t')\vec{g}_{0}(t')dt'. }
By using Eq. (\ref{Mform}) and Eq. (\ref{explicitform}) the above convolution can be expressed as follows:
$\hat{K}\otimes\vec{g}_{0}=-v_{g}(H,\bar{H})^{\rm T}$, where
$H\equiv i\integrale\!\!\!e^{i\delta_{0}(t'-t)}\eta_{z}(v_{g},t'-t)
\left[2|F(t')|^{2}\hat{a}_{0}(t')+F^{2}(t')\hat{a}_{0}^\dagger(t')\right]dt'$. It can be shown that for an arbitrary function $\varphi$ one has
$\integrale\eta_{z}(v_{g},t'-t)\varphi(t')dt' =  \int_{t}^{t-v_{g}^{-1}z}\varphi(t')dt'$.

{\it Squeezing of resonant radiation --} By substituting Eq.(\ref{coupler1}) into Eq. (\ref{rx26}) we have:
\begin{widetext}
\begin{eqnarray}
\hat{a}(z,t)&=&\hat{a}_{0}(t)-iv_{g}e^{-i\delta_{0}t}\left[\int_{t}^{t-z/v_{g}}e^{i\delta_{0}t'}\left\{2|F(t')|^{2}\hat{a}_{0}(t')+F^{2}(t')\hat{a}^{\dagger}_{0}(t')\right\}dt'\right], \label{main1} \\
\hat{a}^{\dagger}(z,t)&=&\hat{a}^{\dagger}_{0}(t)+iv_{g}e^{i\delta_{0}t}\left[\int_{t}^{t-z/v_{g}}e^{-i\delta_{0}t'}\left\{\bar{F}^{2}(t')\hat{a}_{0}(t')+
2|F(t')|^{2}\hat{a}^{\dagger}_{0}(t')\right\}dt'\right], \label{main2}
\end{eqnarray}
\end{widetext}
Equations (\ref{main1}-\ref{main2}) are the main results of this paper. They show that creation and annihilation operators associated with the RR photons become correlated via the classical soliton $F(t)$, even if photons are uncorrelated at $z=0$. This correlation leads to a novel kind of squeezing (which we call {\em hybrid squeezing}), that is conceptually an intermediate case between the SPM-induced squeezing of pulses \cite{blow} and the FWM-induced squeezing of CWs \cite{shelbyfwm}, and is reported for the first time in this paper.

In order to investigate the properties of the hybrid squeezing we construct the two-time radiation noise function $V(z,t,t')$ of the quadrature operator $\hat{X}(z,t) = \hat{a}(z,t)e^{-i\Phi}+\hat{a}^\dagger(z,t)e^{i\Phi}$, where $\Phi$ is a local oscillator phase \cite{blow,foxbook}: $V(z,t,t')\equiv \langle0|\hat{X}(z,t)\hat{X}(z,t')|0\rangle - \langle0|\hat{X}(z,t)|0\rangle\langle0|\hat{X}(z,t')|0\rangle$, where $\ket{0}$ denotes the vacuum state. The full temporal dependence of $V$ must be retained, since the temporal profile of the soliton breaks the translational invariance of $V$ along $t$, analogously to what happens in SPM-induced squeezing \cite{blow}, but unlike the case of FWM-induced squeezing \cite{shelbyfwm}.
Function $V(z,t,t')$ can be written in a semi-analytical (but somewhat cumbersome) form, which we do not show here.

The radiation noise spectrum $S(z,\omega)$ is a {\em real} function, and can be calculated by using the standard formula
\eq{spectrum1}{S(z,\omega)\equiv\frac{1}{T}\int_{-T/2}^{+T/2}dt\int_{-T/2}^{+T/2}dt'\ V(z,t,t')e^{i\omega(t-t')},} where $T$ is the integration time of the spectrum analyzer \cite{blow}. Condition $S=1$ corresponds to shot noise level, and $S(z=0)=1$ at the beginning of propagation. For sufficiently long integration times ($T\gg t_{0}$), expression (\ref{spectrum1}) turns out to be independent of $T$, since $V(z,t,t')$ is a localized function in both $t$ and $t'$ due to the soliton localization itself. Squeezing of quantum noise is achieved in those spectral regions for which $S(\omega)<1$. We now restrict our attention to the resonant frequency $\omega=\delta_{0}$.

\begin{figure}
\includegraphics[width=8cm]{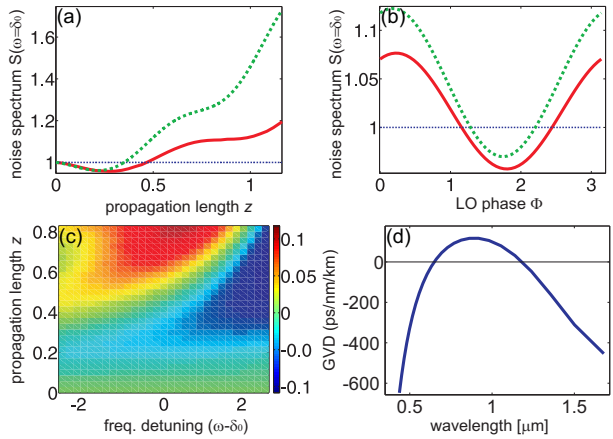}
\caption{(Color online) Squeezing spectrum in various conditions and parameters. (a) Noise spectrum $S(\delta_{0})$ versus propagation distance $z$ for fixed value of the local oscillator phase $\Phi$ = 1.8. (b) Noise spectrum $S(\delta_{0})$ versus $\Phi$ for fixed value of propagation distance $z$ = 0.28. The green dotted and red solid curves in (a,b) correspond, respectively, to the soliton wavenumber $q$ = 0.5 and $q$ = 0.25. The black strait lines in (a,b) represent the shot noise level. (c) Contour plot of ($S$ - 1) showing the $z$-evolution of the noise spectrum at various noise frequency $\omega$. Parameters used in (c): $\Phi$ = 1.8; $q$ = 0.25. Other parameters used in (a,b,c): $\alpha_{3}$ = -0.2; all other HOD coefficients vanish; integration time $T$ = 10; with these parameters the resonant radiation frequency $\delta_{0}$ $\simeq$ -2.7. (d) GVD of the PCF used in this work. At the soliton wavelength ($\lambda_{\rm S}=1.2$ $\mu$m), $\alpha_{3}\simeq-0.2$. \label{fig1}}
\end{figure}

In Fig. \ref{fig1}(a), $S(\omega=\delta_{0})$ as a function of $z$ is plotted with fixed values of $\Phi$ and $\alpha_{3}$, for two different values of the soliton amplitude $q$. This demonstrates the existence of an optimal value of the fiber length for which maximum squeezing of RR is achieved. For large values of $z$, squeezing is lost, so that one would need relatively short pieces of fiber to observe the described phenomenon. This is also advantageous in practice, since in this case the Raman effect, which manifests itself through the Raman self-frequency shift of solitons \cite{RSFS}, does not have time to take place and can be neglected, as we also do here in the equations. Of course in a more precise treatment the effect of the Raman nonlinearity on the correlations must be considered. This analysis will be reported elsewhere.

Fig. \ref{fig1}(b) shows the evolution of $S(\delta_{0})$ versus the local phase oscillator $\Phi$, for fixed values of $z$ and $\alpha_{3}$, plotted for two different values of $q$. In this case, due to the structure of function $V(z,t,t')$, there is a periodic dependency of $S$ on $\Phi$, and maximum squeezing occurs periodically in $\Phi$ with a period equal to $\pi$. In most cases, in our numerical simulations we managed to obtain states with squeezing parameters $S(\delta_{0})$ down to around 0.9, which should be easily detectable in possible future experiments.

It is also interesting to explore what happens to the radiation spectrum around $\omega=\delta_{0}$. In Fig. \ref{fig1}(c) the contour plot of $S(\omega)-1$ versus the propagation distance $z$ is shown. For $\alpha_{3}=-0.2$, the solitonic source of the radiation is located at $\omega-\delta_{0}=2.7$, exactly at the right edge of the spectral window. In the plot, the blue colour indicates those regions for which squeezing is possible [$S(\omega)<1$], while red indicates anti-squeezing [$S(\omega)>1$]. Such figure also shows that, for each given value of $\omega$, there is an optimal value of $z$ that shows maximal squeezing, consistent with Fig. \ref{fig1}(a). Moreover, it is evident from Fig. \ref{fig1}(c) that one has a more pronounced squeezing effect in close proximity of the spectral body of the soliton.

\begin{figure}
\includegraphics[width=0.48\textwidth]{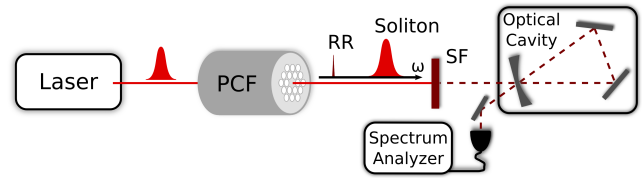}
\caption{(Color online) Schematic of the proposed experimental verification. A pump laser provides optical pulses that propagate through the PCF as solitons. The generated resonant radiation (RR) is isolated with a spectral filter (SF) and coupled into an optical cavity. A photodiode at one cavity output in combination with a spectrum analyzer can be used to measure the noise spectrum around the RR frequency $\delta_0$. Tuning the cavity provides access to the quantum fluctuations at any field quadrature.
\label{fig2}}
\end{figure}

{\em Proposed experimental verification --}
Here we propose an experimental way to investigate the squeezing of RR in PCF. As a source, {\it solid-core} PCF is advantageous, since its GVD can be engineered almost at will by choosing an appropriate hole spacing (pitch) $\Lambda$ and hole size $d$ in the fabrication process. This offers great flexibility in where the RR can be generated and the fiber structure proposed here is merely one of many possible examples. In Fig.~\ref{fig1}(d) we show the GVD and the refractive-index profile of a solid-core PCF with $\Lambda = 1.1~\mu$m and $d=0.8~\mu$m. This fiber exhibits a third-order dispersion coefficient $\alpha_{3}=-0.2$ at the input soliton wavelength $\lambda_s=1.2$ $\mu$m, values that we have used throughout this paper. The resulting RR is generated in the normal dispersion regime at $\lambda_0 \simeq 1.38~\mu$m. To measure the generated squeezed vacuum at frequencies around $\lambda_0$ one can employ usual homodyne techniques, in which the RR is mixed with a strong local oscillator (LO) field that shares the same mode characteristics as the RR. Alternatively, this mode-matching condition can also be met by employing a self-homodyne method based on the dispersive action of optical cavities near resonance~\cite{coelho_science326, villar_amjp76}. In this approach, the carrier of the RR essentially acts as the LO and is interfered with its sideband frequencies via an empty optical cavity. The signal reflected by the cavity is sent to a photodiode whose photocurrent can be studied using a standard spectrum analyzer (Fig.~\ref{fig2}). Note that the analysis frequency corresponds to the ``frequency detuning'' shown in Fig.~\ref{fig1}(c). Quantum fluctuations at any quadrature are simply accessed by scanning the cavity's detuning with respect to $\delta_0$. This causes the noise ellipse to rotate relative to the mean field and thus corresponds to changing the LO phase in a traditional homodyne setup. The use of a properly designed PCF could also help to suppress the guided acoustic wave Brillouin scattering (GAWBS, see Ref. \cite{elser}), which affects the squeezing performance and is neglected here for simplicity. A more precise analysis will include the effect of the Raman and Brillouin scattering in the formalism.

{\em Conclusions --}
By using an original method based on Green's function, we have demonstrated the existence of a novel kind of squeezing, which occurs in correspondence of the resonant quasi-monochromatic radiation emitted by a soliton near the zero-GVD point of a PCF. Correlation between photons occurs at the resonant frequencies and is induced by the presence of the effective potential generated by the soliton. An experimental configuration for observing the presented phenomenon has been proposed. Important applications of this new kind of squeezing include quantum-enhanced measurements, quantum-information processing and quantum cryptography.

This research is funded by the German Max Planck Society for the Advancement of Science (MPG) and was supported by the EC under the grant agreement CORNER (FP7-ICT-213681). K.N.C. acknowledges support from the Alexander von Humboldt Foundation.

\end{document}